# Solid-state batteries enabled by ultra-high-frequency self-heating

Buyi Zhang[1,2], Divya Chalise[1,2], Yuqiang Zeng[2,3], Sumanjeet Kaur[2], Chris Dames[1], and Ravi S. Prasher[1,2]*

[1]Department of Mechanical Engineering, University of California, Berkeley, CA, 94720, USA

[2]Energy Storage and Distributed Resources Division, Lawrence Berkeley National Laboratory, Berkeley, CA, 94720, USA

[3]School of Microelectronics, Southern University of Science and Technology, Shenzhen 518055, China

*Corresponding author: rsprasher@lbl.gov.

**Abstract**

Solid-state batteries (SSBs) are promising next-generation batteries due to their high energy density and enhanced thermal stability and safety. However, their sluggish kinetics and transport at room temperature results in high internal impedance and critically reduces the attainable discharge energy density. Taking advantage of their strong temperature-dependent ionic conductivity, here we introduce ultra-high frequency ($>10^5$ Hz) self-heating (UHFSH) of SSBs, which can rapidly warm up the batteries from room temperature to operating temperature (~65 °C) in less than a minute. As proof of concept, UHFSH experiments were conducted on symmetric solid-state cells with lithium aluminum germanium phosphate (LAGP) electrolyte with different configurations. Using an experimentally validated model, pack-level simulations predict fast heating (50 K/min) and minimized heating energy consumption (less than 4%). Without any modification of the materials or structure of the batteries, our non-intrusive self-heating strategy enables the SSBs to discharge more than two-fold energy in 25 °C ambient.

1. Introduction

Over the past decades, lithium-ion batteries (LIBs) have been widely used as energy storage devices in mobile electronics and electric vehicles (EVs)[1]. However, traditional LIBs using a liquid electrolyte still suffer from relatively lower energy density and significant safety problems due to the flammable nature of the electrolyte. Solid-state batteries (SSBs) stand out as one of the most promising next-generation battery technologies to overcome these issues. By using a lithium-metal anode, SSBs can potentially offer gravimetric and volumetric energy densities that are 40% and 70% higher than those of LIBs, respectively[2]. Moreover, benefiting from their inherent mechanical rigidity and high cationic transference number, inorganic solid electrolytes (ISEs) can potentially improve the fast charging performance and degradation of batteries[3]. Additionally, replacing the organic liquid electrolyte with a non-flammable solid-state electrolyte (SSE) can enhance the thermal stability window and address safety concerns[3–5]. SSBs with ISEs have shown superior cycling performance up to 100 °C,[6] whereas LIBs suffer from accelerated side reactions and degradation at higher temperatures (~60 °C)[7,8]. However, for many SSEs except for a few sulfide electrolytes, at room temperature (RT), the lithium-ion transport kinetics in the electrolyte/electrodes and at the interfaces[9,10] is very sluggish. Sluggish kinetics at RT imply significantly lower ionic conductivity, thereby leading to many problems such as significantly decreased attainable energy/power density of SSBs and the inability to discharge the battery fast enough at RT. The inability to discharge the battery at RT results in the practical challenge that an EV may not start itself. Although some SSBs can discharge enough power at RT, their energy densities are still limited because of low electrode loading and thick electrolytes[11].

To decrease the large internal resistance of SSBs, various heating strategies have been proposed as a mild temperature rise of 50 °C can result in an increase in the ionic conductivity of the SSE by orders of magnitude[9,12–14]. For example, the ionic conductivity of solid polymer electrolytes (SPEs) increases by 1–2 orders of magnitude[9,15]. For ISEs such as lithium lanthanum zirconium oxide (LLZO) and lithium aluminum germanium phosphate (LAGP), it can increase by as much as 5 times, reaching levels comparable to those of liquid electrolytes at RT ($10^{-3}$–$10^{-2}$ S cm$^{-1}$)[2]. Furthermore, with the same temperature rise, the interfacial resistance between electrodes and solid electrolytes is reduced by one order of magnitude[14]. Therefore, in lab environments, additional heating environments such as a temperature chamber and oven are usually used to achieve excellent cycling performance of SSBs[11]. However, external heating strategies such as convective heating are not practical in actual applications such as an EV for many reasons such as the very slow temperature ramp rate due to the high thermal mass of the battery packs and bulkiness of designing a convective heating system itself. Recently, cell-level heating strategies have been developed to circumvent this problem by embedding a thin heater in the cell itself for both LIBs at very low temperatures and SSBs at RT[15–17]. Wang *et al.* proposed to embed nickel foil heaters inside the battery to achieve a heating rate of 1 °C/s,[16] and likewise, Ye *et al.* integrated a thin nickel film in the current-collector layers with a polyimide substrate[15]. However, these approaches are intrusive, raising safety concerns, and require additional materials and processing in the existing battery manufacturing chain[18], which has been refined by the industry over several decades.

An ideal heating strategy would take the best of the external and internal heating approaches, i.e., the strategy would be non-intrusive while providing local heating at the cell level to enable a very high temperature ramp rate. A high temperature ramp rate is very important from a practical viewpoint as this will enable the EV to start quickly rather than waiting for a battery to warm up to the desired temperature. Alternating current (AC) self-heating of a battery is one such non-intrusive method[19]. Depending on the frequency, it completely avoids any structural changes within the battery. AC heating has been applied to traditional LIBs to enable starting the battery at cold temperatures. The temperature ramp rate of AC self-heating for LIBs is ~0.1 °C/s[19], which is much faster than conventional external heating; however, it is still significantly lower for practical use as it takes minutes to reach the ideal ionic conductivity to discharge the battery to start the vehicle.

In this paper, we apply a similar idea to SSBs; however, the temperature ramp rate is nearly one order of magnitude higher than those reported for LIBs, i.e., close to 1 °C/s, which enables discharge of the full capacity at a reasonable C-rate, making it practically viable to start the vehicle within a much shorter time. This high temperature ramp rate is achieved by insulation and ultra-high frequency self-heating (UHFSH), where the frequency is in the MHz range as opposed to the kHz range explored for LIBs in the literature. Without an extra embedded heater or any modification of the internal structure of the cells, our method heats the cell quickly and uniformly, thereby increasing the discharge energy density by two-fold. Additionally, with thermal insulation improved by the pack-level configuration, the heating energy consumption is only 3.5% of the overall battery energy.

## 2. Overview of ultra-high frequency self-heating

Fig. 1a shows the energy discharge density for LIBs with a nickel manganese cobalt (NMC) and lithium iron phosphate (LFP) cathode and an SSB with a LFP cathode as a function of temperature. Both share the characteristic that when the temperature rises, all the kinetic and transport processes are accelerated. For commercial LIBs (citation), at low temperature, their internal impedances are large, leading to a low terminal voltage and reduced discharge capacity. Their discharge energy density approaches the theoretical capacity when discharging at room temperature (~25 °C) and may decrease slightly due to the occurrence of other side reactions at high temperature. However, for SSBs, their discharge energy density performance is continuously enhanced as the temperature increases beyond room temperature[15]. There are a few differences between LIBs and SSBs, which make ultra-high-frequency self-heating (UHFSH) easier and the discharge time much shorter for SSBs. SSBs can have less thermal mass per watt-hour than LIBs (15%–20%, Supplementary Table 1), and SSEs, especially oxide solid electrolytes (LLZTO and LAGP), have higher thermal conductivity (1.4 and 2.2 W m$^{-1}$K) than liquid electrolytes (0.2 W m$^{-1}$K)[20], which imply faster heating speed and smaller temperature gradient.

The mechanism behind UHFSH of SSBs with lithium anodes with alternating current is illustrated in Fig. 1b. During charging (positive current), the lithium ions are moving from cathode material particles to the lithium anode through a solid electrolyte, and during discharging (negative current), they move in the reverse direction. At the electrode/electrolyte interface, both electrochemical reactions and the formation of an interfacial electrical double layer are occurring. The total current $i$ is composed of the reaction (faradaic) current $i_{rxn}$ and interfacial electrical double layer current $i_{inf}$. For DC signals and low-frequency excitations, the interfacial capacitance is large, resulting in no or a negligible interfacial current. Consequently, the total current is nearly equivalent to the reaction current. As the signal frequency increases (Fig. 1c), the interfacial electrical double layers, which act like a capacitor, allow a more significant current to pass through at high frequencies (shown in Fig. 1b). The reaction current goes down, thereby reducing any chemical changes to the SSBs. In addition, the reduced capacitive reactance at high frequency leads to the reduction of the total impedance magnitude, thereby increasing the total current at fixed magnitude of alternating voltage. As a result, the total heating rate increases with increasing frequency (Fig. 1c). However, induction effects come into play when the frequency is too high (discussed later). We can determine the optimal frequency for the largest heat generation, leading to the largest temperature ramp rate based on the electrochemical impedance data of SSBs. An additional constraint applied on UHFSH is the need to limit the voltage to the highest allowable voltage for a given chemistry, which is 4.35–2.35 V for $LiCoO_2/LiNi_xMn_yCo_{1-x-y}O_2|Li$ cell. Fig. 1d summarizes our UHFSH results (discussed later) and provides a comparison with conventional external heating[21–24] and AC heating[17,25,26] for LIBs. UHFSH resulted in an increase of the heating rate by nearly one order of magnitude.

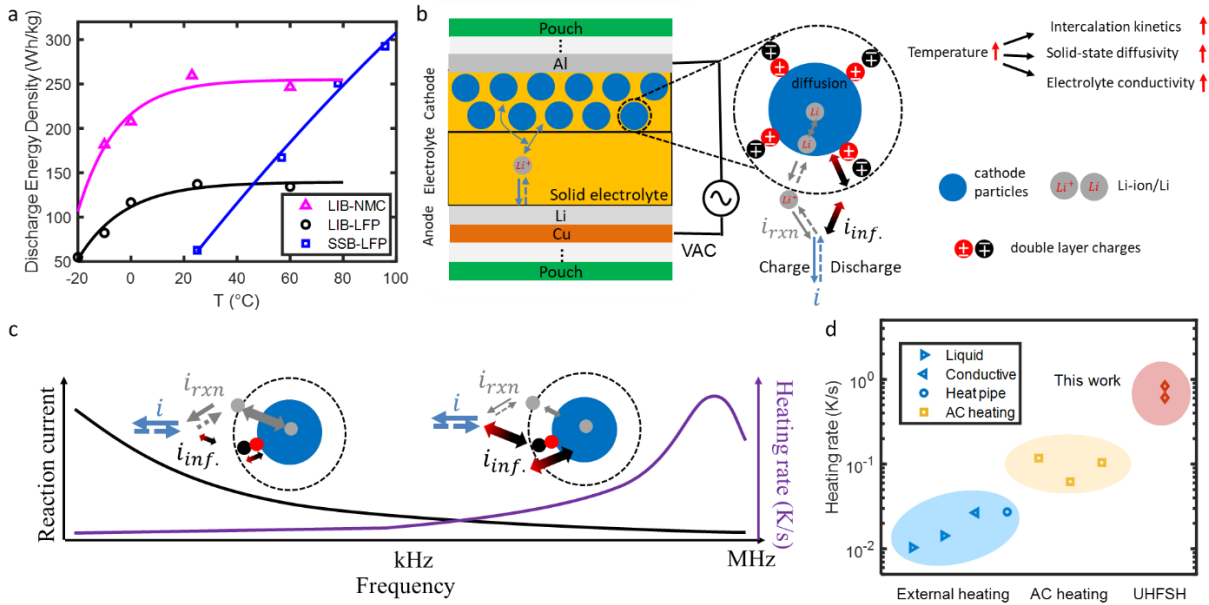

Fig. 1 **a.** Discharge energy density as a function of temperature for LIBs (nickel manganese cobalt (NMC)[27], lithium iron phosphate (LFP)[28]) and solid-state batteries[15] (polymer-based LFP|Li) with

the solid line serving as a guide to the eye. **b.** Schematic of SSBs under certain volts alternating current (VAC). The alternating current $i$ through the cell splits into a capacitive component $i_{inf}$ passing through the electric interfacial capacitance and a reactive component $i_{rxn}$ driving electrode reactions. **c.** Effect of frequency on reaction current and heating rate; the reaction current decreased and the heating rate increased under a constant alternating voltage. **d.** Temperature ramp rate (heating rate) comparison between external heating[21–24], AC heating[17,25,26] ($10^{1-3}$ Hz), and this work ($10^{5-6}$ Hz).

## 3. Ultra-high frequency self-heating experiments and modeling

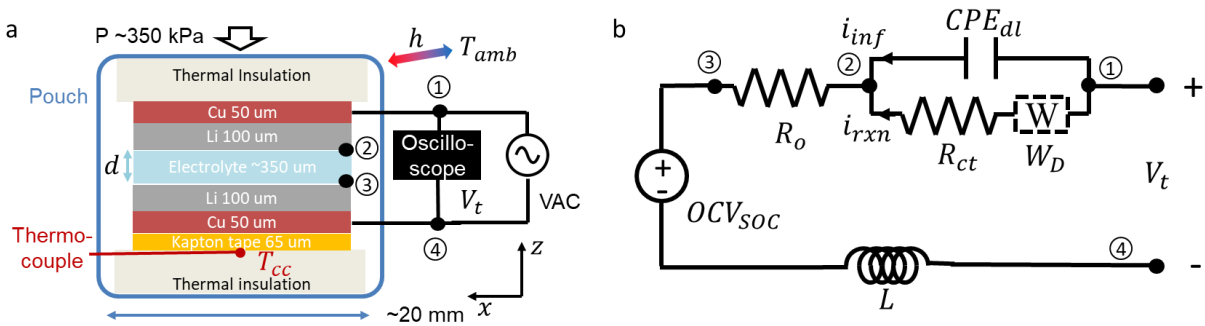

Fig. 2 **a.** Experimental setup for UHFSH of LAGP symmetric cell. **b.** Reduced equivalent circuit model of SSBs

As a proof-of-concept, we conducted UHFSH experiments on a symmetric LAGP cell. These experiments were also used to validate our numerical model using electro-thermal co-simulation in COMSOL. The structure of the built single-layer SSB is shown in Fig. 2a. The LAGP electrolyte (350 μm) is sandwiched by two lithium electrodes (100 μm) and the copper current collectors (50 μm). Electrical connections were made to the copper current collectors by spot welding 50-μm-diameter insulated copper wires. Metal tabs were connected with the other sides of the wires and served as the terminals for the pouch cell. A T-type thermocouple is placed at the center of the copper current collector with Kapton tape (65 um) as the electrical insulation layer in between. The temperature signal from the thermocouple is denoted as $T_{cc}$. The environment temperature is denoted as $T_{amb}$. The cell is stacked with the thermal insulation materials (aerogel) to provide good thermal insulation. The full stack shown in Fig. 2a is vacuum sealed inside a pouch and pressed under ~350 kPa. The high-frequency voltage source is provided by a function generator. An oscilloscope is connected to the cell terminals to record the cell voltage.

The solid-state cell shown in Fig. 3a was first stabilized at room temperature for 10 min and then heated under ±2 V sinusoidal excitation (Fig. 3b). The voltage range was selected as an example and was approximately half the maximum stable potential of LAGP[29]. The overall voltage change (4 V in this case) must be kept below the maximum stable voltage. We swept the frequency from 0.5 to 7 MHz to investigate the frequency dependence. For a single-layer-structured cell (Supplementary Figure 1a), as the frequency increased from 0.5 to 7 MHz, the temperature rise (*ΔT*) increased after 2 min from 0.45 °C to 0.8 °C (Fig. 3c). *ΔT* became stable around 0.85 °C when the excitation frequency was beyond 2 MHz. For stacked double-layer cells (Supplementary Figure 1b), the thermal insulation conditions were better since the ratio of the surface area to volume was reduced. As a result, *ΔT* in 2 min at different signal frequencies was higher compared to that of the single-layer cell, as shown in Fig. 3c. The maximum *ΔT* in 2 min was approximately 1 °C. Moreover, the steady-state *ΔT* for the double-layer was higher than that for the single-layer cell, as shown in Fig. 3d. Furthermore, we conducted electro-thermal co-simulation in COMSOL numerical simulation to model the heating behavior of solid-state cells using the UHFSH method. We built the equivalent circuit model for our symmetric cell. The lithium anode and electrode/electrolyte interface impedance were modeled as the charge-transfer resistance in series with a Warburg element and in parallel with electrical double layer capacitance. The internal ohmic resistance mainly originates from the solid electrolyte. For the symmetric LAGP cells we constructed, because the lithium diffusion for the lithium anode is negligible, the Warburg element was removed. In large-capacity cells, inductance of the layer may become important (discussed later); however, for these small single- and double-layer cells, they are negligible.

One simulated cross-section temperature distribution is shown in Supplementary Figure 2. The thermal properties such as the heat capacity and thermal conductivity of the LAGP electrolyte, lithium foil, and other materials were measured using differential scanning calorimetry and the 3-omega method or taken from the literature (Supplementary Table 3). The heat generation is calculated from cell-impedance data and voltage using $P = V_{cell}^2 Z_{Re}/|Z|^2$. $V_{cell}$ is the cell terminal voltage, $|Z|$ is the magnitude of the cell impedance, and $Z_{Re}$ is the real part of the cell impedance. In the model, the only fitting parameter is the heat-transfer coefficient ($h$) between the pouch and the ambient (Fig. 2a). The estimated $h$ for the single-layer cell and two-layered cell are 25 and 20 $W/(m^2 K)$, respectively. The simulated temperature evolutions of the pouch cells match well with the experimental data, as shown in Fig. 3c and 3d. We use this model to predict large-scale cell-level and pack-level performance of SSBs, as discussed in the next section.

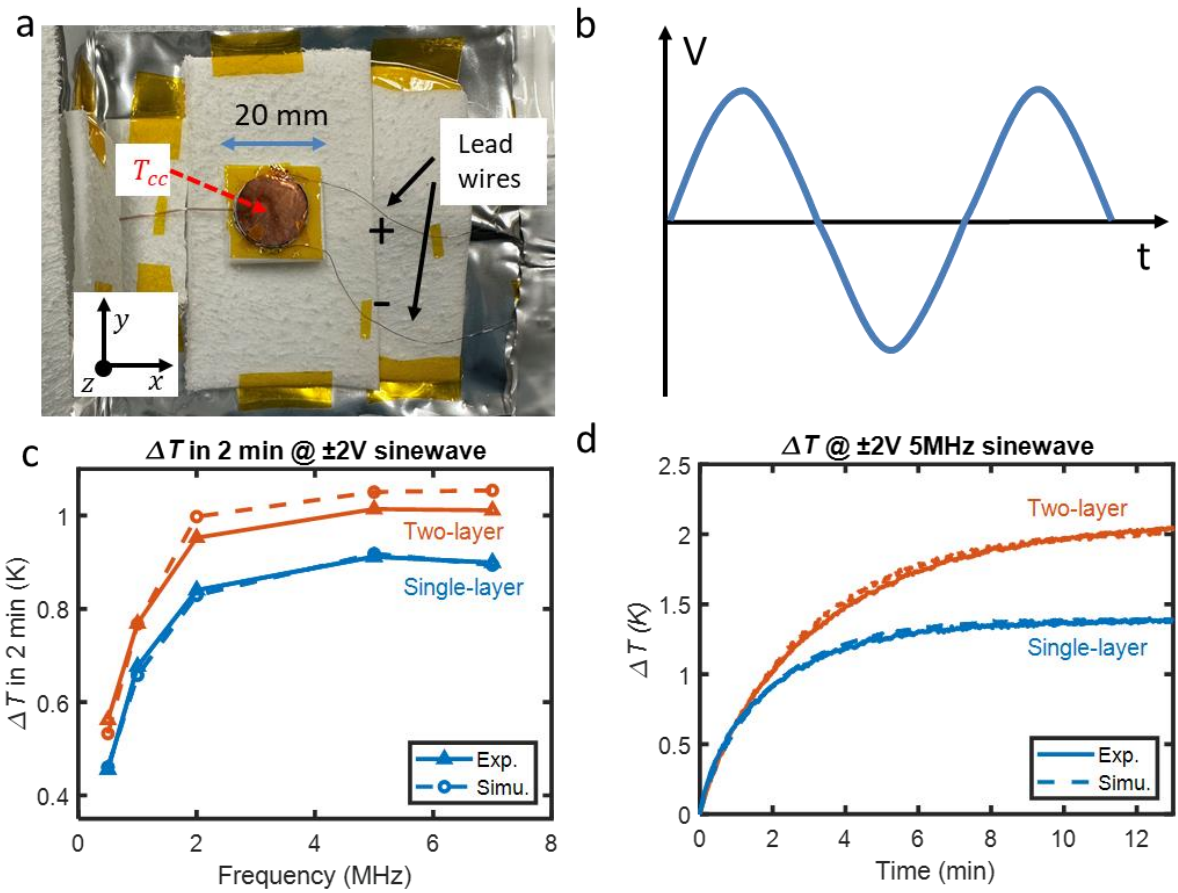

Fig. 3 **a.** Images of the assembled single-layer symmetric LAGP pouch cell (12-mm diameter 100-μm Li electrode, 23.3 mAh) and **b.** sinusoidal voltage wave used in experiments. Schematics of single-layer and two-layer cells are provided in Supplementary Fig. 1. **c.** Temperature rise $\Delta T = T_{cc} - T_{amb}$ ($T_{cc}$ is measured at the cell current-collector center, as shown in Fig. 2a; $T_{amb}$ is the ambient temperature) in 2 min across different AC pulsed voltage frequencies for a single-layer cell and stacked two-layer cell. **d.** Temperature evolution for the single-layer cell and stacked two-layer cell as a function of time.

## 4. Heating performance on cell and pack level

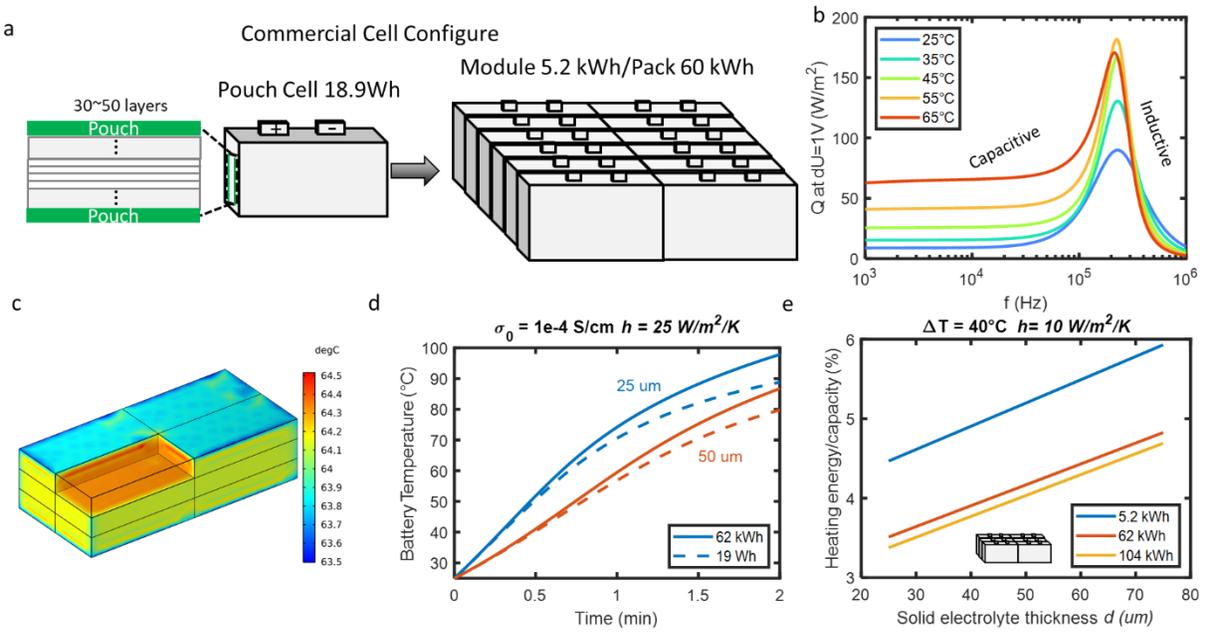

Fig. 4 **a.** Battery configurations for different capacities (small to large): single layer to pouch cell and module and pack level. **b.** AC heating power per electrode area as a function of frequency under ±1 V excitation at various uniform cell temperatures. The overall voltage change of 2 V is within the stable voltage of LCO|Li. **c.** Simulated temperature distribution of pack (62-kWh pack after 45 s heating). **d.** Temperature evolution for 19-Wh pouch cell and 62-kWh pack with 25-um and 50-um electrolyte thickness under ±1-V pulse at optimal frequency of 0.2 MHz shown in Fig. 4b, assuming ionic conductivity at 25°C, $\sigma_0=10^{-4}$ S/cm, h=25 W/m²/K. **e.** Heating energy consumption of different SSB sizes and solid-electrolyte thickness.

In this section, we use our validated electro-thermal co-simulation model to predict the performance of SSBs at the pouch and pack level to understand the impact of UHFSH at the commercial scale. Figure 4a shows the multilayer configurations of commercial batteries from the cell to pack level. In general, a cell is composed of many (30–50) stacks of layers for a pouch cell or a jelly roll in a cylindrical cell. Moreover, in EV applications, hundreds of cells are arranged in modules. They are connected in series or parallel to achieve the desired voltage and energy capacity. A whole battery pack could consist of several to tens of modules. One of the biggest differences between the lab-scale single- and double-layer systems vs. a commercial-scale system is that the surface to volume ratio is much smaller for the commercial-scale systems. This smaller ratio greatly reduces the heat loss to the environment, thereby enabling a faster temperature rise. Another difference is that large multilayer cells have more inductance in the cell impedance[30–33]. The inductance behavior (negative imaginary impedance) is trivial for small-capacity cells; however, it must be considered for large-capacity batteries because of their multilayer structure.

**4a Effect of voltage frequency on heating power**

Figure 4b shows the estimated heat generation per electrode area using our equivalent circuit model ($P = V^2 Z_{Re}/|Z|^2$, Supplementary Note 1) under VAC as a function of temperature and input frequency with 400 um thick electrolyte. Considering the full-cell LCO|Li voltage range (4.35–2.35 V), we are using ±1 V amplitude and a square wave form to maximum the heat generation. The heating power as a function of frequency is not monotonic, unlike the single-layer cell case. At low frequency, the large interface and charge-transport impedance dominates and leads to low heat generation. At the higher frequency, the electrical double layer behaves like a capacitor, resulting in shorting of the charge-transfer resistance. Therefore, the total impedance, both the imaginary and real part, is reduced, resulting in significant increase in heat generation. However, if the frequency is too high, the inductance becomes dominant, and the overall impedance increases, resulting in lower heat generation.

**4b Effect of solid-state electrolyte on heating rate**

Both the ionic conductivity and thickness of the solid-state electrolyte are very critical as they are part of the overall electrical impedance of the cell. The broader research and development (R&D) community and the industrial community are actively working on reducing the thickness of solid-state electrolytes and increasing their ionic conductivity[11,34]. We investigated the impact of UHFSH for a wide range of ionic conductivity and thickness of the electrolyte. Using the calculated heat generation from the last section, we started the simulation on a 19-Wh LCO|LAGP|Li pouch cell that consists of tens of layers (Supplementary Figure 3a). We applied a ±1-V square wave signal with the optimal frequency 0.2 MHz (Fig. 4b) to estimate the temperature rise. With a fixed 25-μm electrolyte thickness, the same as the separator thickness in LIBs, we show how the heating time would decline as the electrolyte ionic conductivity increases in Supplementary Figure 3b. A contour plot of time to increase $\Delta T$ by 40 °C as a function of electrolyte thickness and ionic conductivity is presented in Supplementary Figure 3c. With smaller thickness and higher ionic conductivity, the required heating time decreases and becomes sub-minute.

To reproduce the thermal behavior in compact battery packs of EVs, besides the cell level, we built three different battery sizes: 5.2 kWh (module), 62 kWh (small pack), and 104 kWh (large pack). Figure 4d shows that for a 25-μm-thick electrolyte, the battery (pack) temperature can increase to over 70 °C in 1 min. The temperature rises at an optimal frequency in 2 min for a 62-kWh battery pack are 10 °C higher than that of a 19-Wh single cell. The small surface to volume ratio of the pack enhanced the battery insulation and improved the heating rate.

**4c Heating energy consumption**

Figure 4e shows the percentage of heating energy consumption over the energy stored in the battery using our electrothermal model of SSBs for the whole pack size, with comparison to the module level. The variation across different solid-electrolyte thicknesses are also shown. In terms of raising the battery by 40 °C up for 1-h operation with h=10 W/m$^2$K, the energy required to heat a 104-kWh battery pack is only 3.4% of the total battery energy. Additionally, the reaction heat released during battery discharge is not included in the calculated energy consumption, which will further reduce the energy loss to maintain the temperature during operation. Based on our study of UHFSH on SSBs so far, this fast heating method could be further optimized in a practical solid-state battery management system to achieve higher obtainable energy density, better cycling, and fast start-up performance.

**Conclusions**

In summary, we have successfully developed an ultrafast-heating method for SSBs using an ultra-high-frequency voltage pulse, which can enable SSBs to start from RT in minutes. In contrast to previous advances in battery materials and interfaces to commercialize SSBs, our method retains the original battery structure and leverages the small heat-energy consumption to achieve a large energy density increase. Considering the new challenges and opportunities for SSBs, their optimal operating temperature can be adjusted promptly via the UHFSH in a smart battery thermal management system.

**Methods**

**Electrochemical thermal simulation and verification.** We used the equivalent circuit model and Heat Transfer Module in COMSOL Multiphysics 5.6 for the simulation of SSB heating under different alternating voltage excitation and thermal conditions.

**Solid-state electrolyte preparation.** The LAGP pellets were made from the commercially available sodium super ionic conductor (NASICON) LAGP powder (300-500 nm MSE Supplies). The powders were pressed in a 15-mm diameter stainless steel die under 15000 lb force for 10 min. Then, the pellets were sintered on alumina inside the furnace in air at 700 °C for 10 h. In addition, the pellets were sputtered with 40-nm gold on both sides to enhance the interface contact with the lithium foil.

**SSB with thermal insulation preparation.** With the prepared electrolytes, 100-um-thick lithium foil (MSE Supplies) was punched into 12-mm-diameter coins as the electrodes, and they were cleaned on both sides with a tweezer to remove the surface oxides or contaminants. Then, the cleaned lithium discs were pressed on both sides of the LAGP pellets. The structure was then sandwiched between two 50-um-thick copper (McMaster-Carr) circles with 15-mm diameter. The 50-um-diameter insulated copper wires were spot welded with the copper current collector at one end and with nickel tabs at the other end. We used the wires for better thermal insulation. Then, the nickel tabs served as the battery's terminal and were connected to alternating signals. A thermocouple was placed at the center of one copper circle with the Kapton tape in between as the electrical insulator. The entire structure was sandwiched between the 3.2-mm-thick aerogel blocks (Airloy x103®) with 20-mm squared surface. The cell was finally sealed in a pouch-cell configuration (MTI Corporation) in an Ar-filled glove box ($O_2 < 0.1$ ppm and $H_2O < 0.1$ ppm). After cell assembly, the cell was annealed in a furnace at approximately 65 °C for 7 h.

**UHFSH experiments.** The pouch cell was held with a lab-made pressure jig and load sensor and placed inside the test equity model TEC1 thermoelectric temperature chamber. The cell and environment temperature were recorded with T-type thermocouples (Omega Engineering) by a Pico TC-08 thermocouple data logger. The high-frequency alternating voltage wave was supplied by a RIGOL DG4062 function waveform generator. The cell terminal voltage was measured using an Agilent DSO3202A digital storage oscilloscope. The temperature rises and terminal voltages were documented under various high-frequency sine signals.

**Electrochemical impedance spectroscopy.** The electrochemical impedance spectroscopy (EIS) measurements were performed using a Biologic VMP3 multichannel potentiostat. The symmetric cells were rested for 15 min before tests and under 350 kPa during tests. A signal with 2-V amplitude was applied to the cell with a frequency ranging from 0.1 Hz to 7 MHz with 20 data points per decade of frequency. An equivalent circuit model[35] was used to fit the EIS results (Supplementary Note and Supplementary Figure 4). The parameters determined from the EIS analysis are summarized in Supplementary Table 2.

**Data availability**
The data supporting the findings of this study are available from the corresponding author on reasonable request.

**Code availability**
The code used in this study are available from the corresponding author on reasonable request.

**References**
1. Schmuch, R., Wagner, R., Hörpel, G., Placke, T. & Winter, M. Performance and cost of materials for lithium-based rechargeable automotive batteries. *Nat Energy* **3**, 267–278 (2018).

**Acknowledgments**

The authors would like to thank Dr. Yanbao Fu for assistance with the cell-assembly procedure, Dr. Mike Tucker for providing access to the Biologic potentiostat, and Dr. Kenneth Higa (LBL) for the fixturing clamp setup to apply pressure. The authors acknowledge the support from the Energy Efficiency and Renewable Energy, Vehicle Technologies Program, of the US Department of Energy under contract no. DEAC0205CH11231.

**Author Contributions**

B. Z. and R. S. P. conceived the idea. B. Z. designed and conducted the electrochemical simulations and cell assembly and heating experiments. Y. Z. and D. C. discussed the results. B. Z., C. D., S. K., and R. S. P. contributed to writing of the manuscript.

**Competing Interests**

The authors declare no competing interests.

**Figure Captions**

**Supplementary Information**

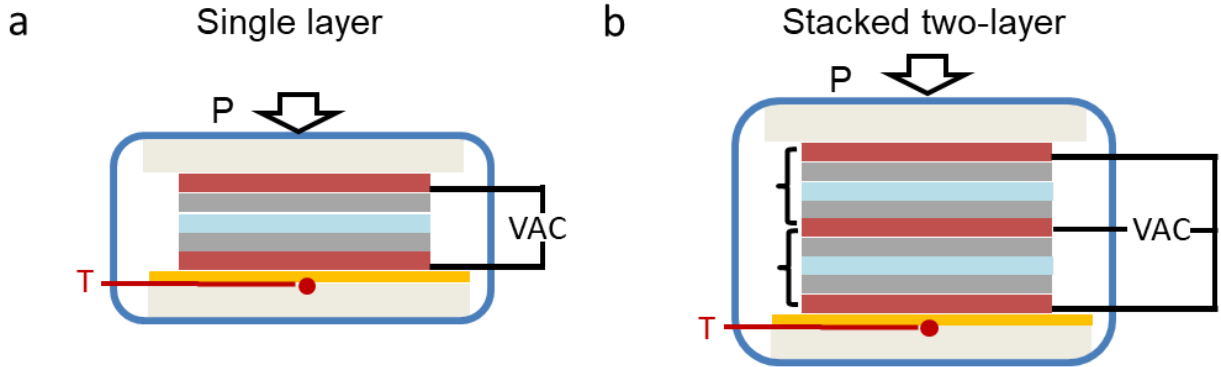

Fig. S1 Schematics of LAGP symmetric cell **a.** Base single-layer cell, **b.** Stacked two-layer cell

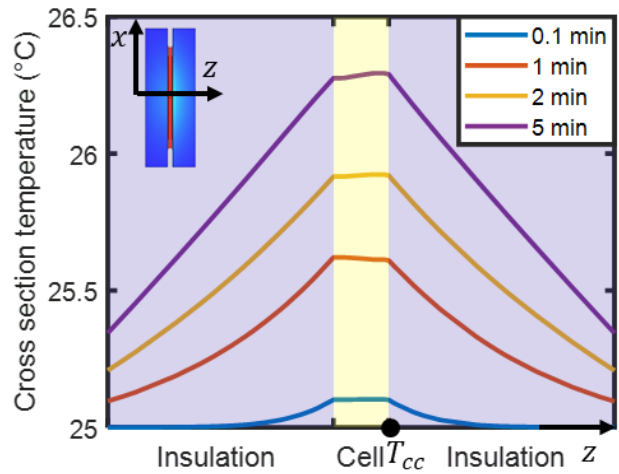

Fig. S2 Simulation results showing the cross-section temperature distribution $T(0,0,z,t)$ of the single-layer solid-state cell under ±2V 5 MHz sinewave at different times using COMSOL.

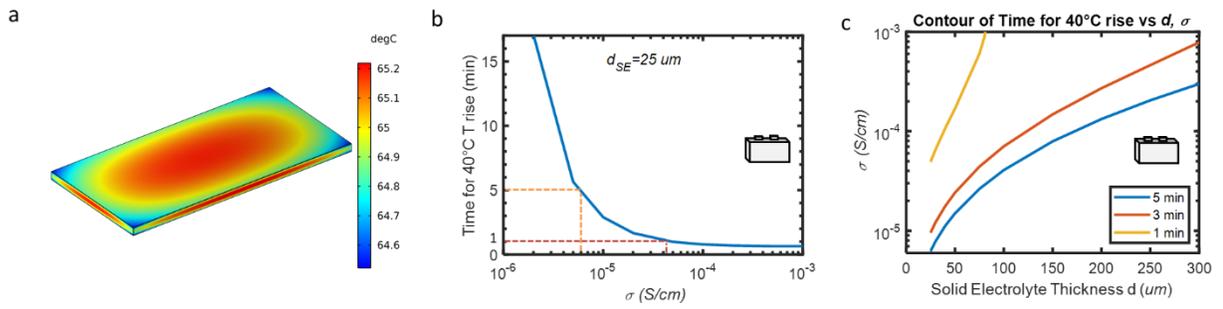

Fig. S3 For a 18.9-Wh (5 Ah) LCO|LAGP|Li cell configuration (10 cm × 5 cm), **a.** Simulation example snapshot for cell with 25-μm thick electrolyte after 52-s heating. **b.** Time required to reach 40 °C rise as a function of ionic conductivity for electrolyte thickness of 25 μm. **c.** Time required to reach 40 °C rise for different solid-state electrolyte thickness and ionic conductivity.

A higher ionic conductivity of the electrolyte results in higher optimal heating power and less heating time. The cell temperature could be raised by 40 °C within 5, 3, and 1 min when the ionic conductivity was $0.5 \times 10^{-5}$, $10^{-5}$, and $10^{-4}$ S/cm, respectively. In addition, the reduction of the electrolyte thickness is more important for fast heating because it will decrease the high-frequency resistance and thermal mass of the cell.
24

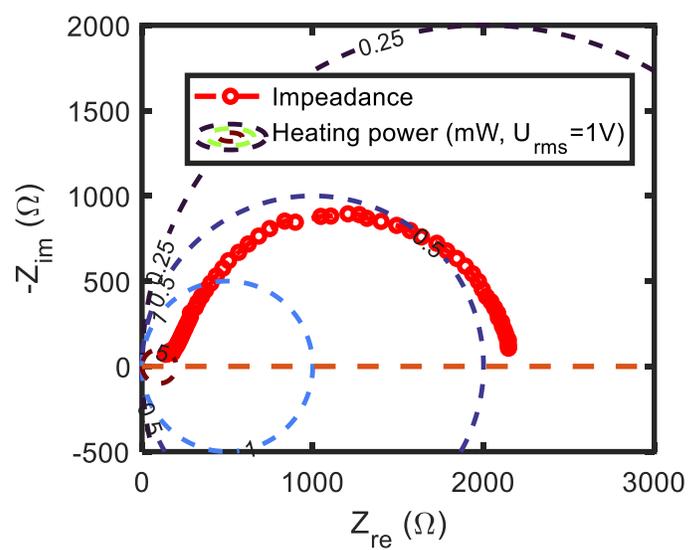

Fig. S4 Electrochemical impedance spectroscopy of LAGP symmetric cells and contour of the heating power under ±1-V pulse of this LAGP cell

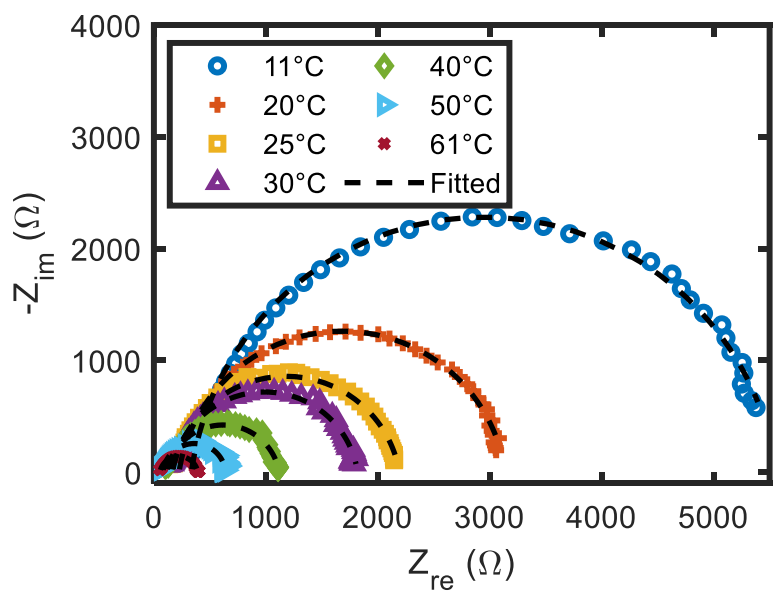

Fig. S5 Electrochemical impedance of LAGP symmetric cells at different temperatures and modeling fitting results

**Supplementary Table 1.** Parameters of cell components to calculate the heat capacity per watthour or area

| Materials | PEO | LAGP | Li | Cu | graphite | separator | liquid electrolyte | LCO | Al |
|---|---|---|---|---|---|---|---|---|---|
| Thickness (μm) | 25 | 25 | 15 | 10 | 70 | 25 | | 71 | 20 |
| Porosity | | | | | 0.35 | 0.41 | | 0.35 | |
| Density (kg/m^3) | 1200 | 3270 | 534 | 8938 | 2240 | 913[36] | 1220 | 5160 | 2699 |
| Specific heat (J/kg/K) | 1690[37] | 698.5 | 3559 | 377 | 750 | 2480[36] | 1301[38] | 744[39] | 921 |

With the same loading of cathode (LiCoO$_2$), the area of the cell is equivalent to the cell's energy. The thermal mass of SSBs and LIBs are calculated

| | PEO | LAGP | Liquid electrolyte |
|---|---|---|---|
| Area heat capacity (J/K/m$^2$) | 417 | 403 | 488 |
| Decreased percentage | 14.61% | 17.46% | 0 |

**Supplementary Table 2.** Parameters and root mean squared error (RMSE) between the experimental data (markers) and fitting results (dashed lines) from the EIS analysis of symmetric LAGP pouch cells in Supplementary Figure 5.

| $T$ (°C) | 11 | 20.6 | 25 | 30 | 40 | 50 | 60.9 |
|---|---|---|---|---|---|---|---|
| $R_0$ (Ω) | 335.0720 | 224.6236 | 167.2584 | 158.9276 | 116.0164 | 81.4454 | 54.1934 |
| $R_1$ (Ω) | 5225.6 | 2915 | 2025.6 | 1660.6 | 1003.6 | 558.6327 | 337.9970 |
| $Q$ ($s^n$/Ω) | 3.7303×10$^{-9}$ | 4.2689×10$^{-9}$ | 4.9972×10$^{-9}$ | 4.4739×10$^{-9}$ | 5.7413×10$^{-9}$ | 2.8612×10$^{-9}$ | 6.9431×10$^{-9}$ |
| $n$ | 0.9142 | 0.9086 | 0.8951 | 0.9081 | 0.8914 | 0.9470 | 0.8923 |
| $RMSE$ (Ω) | 53.8049 | 28.1856 | 20.6882 | 19.6008 | 14.2726 | 15.2645 | 3.5717 |

Then, the fitted activation energy $E_a$ for $R_1$ is 43.5 kJ/mol.

**Supplementary Table 3.** Thermophysical properties used in single-layer and double-layer experiments

| Material | Density (kg/m3) | Thermal conductivity (W/m/K) | Specific heat (J/kg/K) | Thickness (mm) |
|---|---|---|---|---|
| LAGP | 2273[a] | 1.8[20] | 808[c] | 0.33~0.42 |
| Kapton tape | 1276[a] | 0.18 cross plane, 0.45 in plane [b] | 1497[c] | 0.0635 |
| Aerogel (Airloy x103®) | 200[a] | 0.031[40] | 2000[40] | 3.2 |

[a]: calculated from mass and dimension measurements

[b]: 3-omega experiments

[c]: differential scanning calorimetry experiments

**Supplementary Note 1. Equivalent Circuit Model for Cell/Pack-Level Simulation**

Figure 2b shows the equivalent circuit model used for fitting the electrochemical impedance spectra: $R_o$ and $R_{ct}$ are the ohmic resistance and charge-transfer resistance, respectively. $CPE_{dl}$ and $W_D$ represent the double-layer capacitance and the Warburg diffusion element, respectively. The first component is an inductance, which comes from the stacks of current collectors and connecting wires.
$Z_L = iwL.$
The second term is the ohmic resistance of the battery, mainly comprised of the electrolyte, electrode, and contact resistances.
$Z_0 = R_0.$
The third component stands for the charge-transfer process at the active material surfaces:
$Z_{ct} = (R_{ct} + Z_w) * CPE_{dl}/(R_{ct} + Z_w + CPE_{dl}),$
where $CPE_{dl} = 1/(iwC_{dl})$ and the lithium diffusion impedance $Z_w = 1/(C_w(iw)^{0.5}).$
The total impedance is the sum of the three components. The parameters used in the model are summarized in the following table. They are hosen based on experiments and literature values.

| L (nH) | $R_0$ ($\Omega*cm^2$) | $R_{ct}$ ($\Omega*cm^2$) | $C_{dl}$ (F/cm²) | $C_w$ (F/cm²) |
|---|---|---|---|---|
| 30[32,33] | $t_{LCO}/\sigma_{LCO} + t_{SE}/\sigma_{SE}$ | 1000×exp($E_a$/R*(1/$T_0$-1/T) | 10⁻⁸ [41] | 10⁻³ [42] |
| $t_{LCO}$ (μm) | $\sigma_{LCO}$ (S/cm) | $\sigma_{SE,LAGP}$ (S/cm) | | |

| | 40 | COMSOL Material Library | $4\times10^{-4}\times\exp(36[kJ/mol]/R*(1/T_0-1/T))$[43] | | |

$E_a = 43.5\ kJ/mol$ is activation energy for charge transfer resistance $R_{ct}$, fitted from Supplementary Table 2, R = 8.3145 *J/mol/K* is the gas constant. $T_0$ = 298.15 *K* is room temperature.

**Supplementary Note 2. Thermophysical Properties for Cell/Pack-Level Simulation**

To simplify the model, the entire pouch cell was treated as a block with effective thermal properties and a volumetric heat source. We determined the effective thermal properties such as the thermal conductivity and heat capacity using a homogenization method based on the multi-layer. The heat-transfer coefficient between the surface and the environment is assumed to be 10 $W/(m^2 K)$.

| Material | Thickness (μm) | Density (kg/m3) | Thermal conductivity (W/m/K) | Specific heat (J/kg/K) |
|---|---|---|---|---|
| $LiCoO_2$ | 40 | 5160[44] | 2.49 cross plane, 21.75 in plane[45] | 695.4+ 2.104×T- 5.356×T²×10⁻³ [39] |
| LAGP | 25~75 | 3270[46] | 20 | 20 |
| Pouch[15] | 105 | 1580 | 0.2 | 1030 |
| Lithium | 15 | | | |
| Al | 15 | COMSOL Library | | |
| Cu | 9 | | | |